# Do bibliometrics introduce gender, institutional or interdisciplinary biases into research evaluations?

Mike Thelwall
Statistical Cybermetrics and Research Evaluation Group, University of Wolverhampton, UK.
https://orcid.org/0000-0001-6065-205X m.thelwall@wlv.ac.uk

Kayvan Kousha
Statistical Cybermetrics and Research Evaluation Group, University of Wolverhampton, UK.
https://orcid.org/0000-0003-4827-971X k.kousha@wlv.ac.uk

Emma Stuart
Statistical Cybermetrics and Research Evaluation Group, University of Wolverhampton, UK.
https://orcid.org/0000-0003-4807-7659 emma.stuart@wlv.ac.uk

Meiko Makita
Statistical Cybermetrics and Research Evaluation Group, University of Wolverhampton, UK.
https://orcid.org/0000-0002-2284-0161 meikomakita@wlv.ac.uk

Mahshid Abdoli
Statistical Cybermetrics and Research Evaluation Group, University of Wolverhampton, UK.
https://orcid.org/0000-0001-9251-5391 m.abdoli@wlv.ac.uk

Paul Wilson
Statistical Cybermetrics and Research Evaluation Group, University of Wolverhampton, UK.
https://orcid.org/0000-0002-1265-543X pauljwilson@wlv.ac.uk

Jonathan Levitt
Statistical Cybermetrics and Research Evaluation Group, University of Wolverhampton, UK.
https://orcid.org/0000-0002-4386-3813 j.m.levitt@wlv.ac.uk

Systematic evaluations of publicly funded research typically employ a combination of bibliometrics and peer review, but it is not known whether the bibliometric component introduces biases. This article compares three alternative mechanisms for scoring 73,612 UK Research Excellence Framework (REF) journal articles from all 34 field-based Units of Assessment (UoAs) 2014-17: peer review, field normalised citations, and journal average field normalised citation impact. All three were standardised into a four-point scale. The results suggest that in almost all academic fields, bibliometric scoring can disadvantage departments publishing high quality research, with the main exception of article citation rates in chemistry. Thus, introducing journal or article level citation information into peer review exercises may have a regression to the mean effect. Bibliometric scoring slightly advantaged women compared to men, but this varied between UoAs and was most evident in the physical sciences, engineering, and social sciences. In contrast, interdisciplinary research gained from bibliometric scoring in about half of the UoAs, but relatively substantially in two. In conclusion, out of the three potential source of bias examined, the most serious seems to be the tendency for bibliometric scores to work against high quality departments, assuming that the peer review scores are correct. This is almost a paradox: although high quality departments tend to get the highest bibliometric scores, bibliometrics conceal the full extent of departmental quality advantages. This should be considered when using bibliometrics or bibliometric informed peer review.
**Keywords**: Research bias; gender; peer review; REF2021; interdisciplinarity

# Introduction

Many countries now employ systematic assessments of publicly funded research institutions to evaluate their performance and/or to allocate performance-based funding (Sivertsen, 2017). These may be carried out primarily by peer review, by peer review informed by bibliometrics, or primarily by bibliometrics, and/or other indicators (Sivertsen, 2017). For example, the UK Research Excellence Framework (REF) is a peer review exercise with bibliometrics supporting 11 of its 34 field-based Units of Assessment (UoAs) and other indicators supporting peer review of research environments. In contrast, Sweden allocates funding based on bibliometric and other indicators, for transparency, reserving peer review for formative research evaluations (Sivertsen, 2017). The potential biases from peer review or indicators are important concerns both formative evaluations and performance-based research funding. For example, if the outcomes are unfair for reasons of institutional focus, gender, or research type (Lee et al., 2013), then the assessment could be inefficient and/or unethical. Despite this important concern, little is known about the biases introduced by peer review or bibliometrics into national research assessments.

This article investigates three important types of potential bias for peer review or bibliometrics: institutional performance, gender, and interdisciplinary research. The first two are general concerns for the efficiency of an assessment system and focus on whether the average quality of institutional scores are related to whether they gain or lose from bibliometrics. The third is both an ethical issue and an efficiency concern if half of all researchers are devalued. The fourth focuses on a type of research that is both difficult to evaluate and widely encouraged in the belief of its scientific and societal value. The following research questions drive this study.
- RQ1: Do grades based on **article-level** citation-based indicators give systematically different **institutional** outcomes than grades based on peer review in any fields?
- RQ2: Do grades based on **journal-level** citation-based indicators give systematically different **institutional** outcomes than grades based on peer review in any fields?
- RQ3: Do grades based on bibliometrics favour **female** researchers compared to grades based on peer review in any or all fields?
- RQ4: Do grades based on bibliometrics favour **interdisciplinary** research compared to grades based on peer review in any or all fields?

# Background

## *Article-level citation-based indicators and research quality*

The core rationale behind using citation counts as an indicator of the value, quality or impact of an academic article is that scientists cite to acknowledge prior influences so that an article's citation count reflects its influence on future research (Merton, 1973). The many flaws of this argument include citations being used for many other purposes, including comparisons, and being influenced by social factors (Lyu et al., 2021). In addition, citations may rarely play a core role in less hierarchical subjects (Lin, 2018). More fundamentally, research quality is generally thought to encompass three dimensions: rigour, significance, and originality (Langfeldt et al., 2020). Of these, citations probably reflect significance most and it is not clear that they are good indicators of rigour and originality (Aksnes et al., 2019). Moreover, citations do not reflect societal impacts (van Driel et al., 2007). Thus, even from a theoretical perspective, it seems unlikely that citation counts closely correlate with research quality within any field, unless its three dimensions usually coincide for some reason or if societal impact, rigour and originality influence all citing behaviours.

From an empirical perspective, several studies have compared scientific impact with peer review quality judgements of academic articles to assess whether citation counts could be a good indicator. For example, a close to zero correlation was found between citation counts and expert ratings of articles in a medical journal (West & McIlwaine, 2002), between most details of methods reporting (i.e., related to rigour) and citation counts for four psychology journals (Nieminen et al., 2006) and for dementia biomarker studies (Mackinnon et al., 2018). By far the largest scale comparison used four-level peer review quality rating REF scores in 36 UoAs for 19,580 journal articles from 2008, finding zero or negative Spearman correlations in five UoAs: Theology and Religious Studies (-0.2), Classics; Philosophy (-0.1), Art and Design: History, Practice and Theory; Music, Dance, Drama and Performing Arts (0). The remaining correlations were all positive and at least weak, with the strongest in Biological Sciences, Chemistry and Physics (all 0.6) and Clinical Medicine (0.7) (HEFCE, 2015, Table A3). Similar correlations were found between quality ratings and the field normalised citation indicator FWCI (Field Weighted Citation Impact), although the disciplinary differences were less extreme. Spearman correlation strengths were zero or negative in four UoAs: Classics; Music, Dance, Drama and Performing Arts; English Language and Literature; Theology and Religious Studies (-0.1). The strongest correlations were in Clinical Medicine and Physics (both 0.6) (HEFCE, 2015, Table A8). Thus, citation counts, whether field/year normalised or not, are imperfect indicators of journal article quality in most academic fields but their value varies greatly between fields, and they are useless in some arts and humanities fields.

Studies that have correlated average citations with average quality scores aggregated at the departmental level have tended to find positive correlations varying in strength from 0.2 to 0.8 (Abramo et al, 2011; Baccini & De Nicolao, 2016; Franceschet & Costantini, 2011; Pride & Knoth, 2018; Rinia et al., 1998; van Raan, 2006). Most previous investigations of the relationship between departmental average numbers of citations and RAE/REF scores also found statistically significant positive correlations, although there were disciplinary differences (e.g., Mahdi, D'Este & Neely, 2008; Jump, 2015). These reveal little about article-level correlations, however, since correlation coefficients naturally increase when data is aggregated (van Raan, 2004).

### *Journal-level citation-based indicators and research quality*

The Journal Impact Factor (JIF) and other average citation impact indicators for journals are widely used informally for formal and informal research evaluations. At the informal level, academic appointment committees lacking the time to read the candidates' papers might use JIFs to help make quick decisions about the quality of research described in a CV (McKiernan et al., 2019). Individual researchers may also consult JIFs when deciding where to publish (Beshyah, 2019; Sønderstrup-Andersen & Sønderstrup-Andersen, 2008). More formally, some national evaluation systems reward scholars for publishing in journals meeting a JIF threshold or include JIFs in performance-based funding formulae (Sivertsen, 2017), although many countries construct bespoke stratified lists of journals to assess or reward research (Pölönen et al., 2021).

In research evaluation contexts, JIFs have the advantage of being relatively transparent, compared to informal ideas of journal prestige shared within a research community. In fields where citation counts are reasonable indicators of research quality, journals with more citations per article would tend to publish better articles, so journal citation rate calculations would give (imperfect) indicators of research quality. They may also be better indicators of the quality of an article than the article's citations in some fields (Waltman & Traag, 2020). Moreover, in fields where JIFs are well regarded, competition to publish in higher-JIF journals would form a positive feedback loop (Drivas & Kremmydas, 2020) in which higher JIF journals increasingly monopolise research that the field regards as high quality. Nevertheless, highly original research may tend to be published in journals with lower impact factors (Wang et al., 2017), so the novelty quality dimension may be captured poorly by journal metrics.

There are many disadvantages of using JIFs for research evaluation that have led to the San Francisco Declaration on Research Assessment (DORA) campaign against them. JIFs have most of the disadvantages of citation counts, as discussed above. For example, in fields where citations are not indicators of research quality, such as much of the arts and humanities, they may be irrelevant (Fuchs, 2014). They are often inappropriately compared between fields, despite large natural variations in field citation rates. There are many technical problems, such as failure to deal appropriately with the skewed nature of citation counts in most, calculation errors, and discrepancies between the numerator and denominator in calculations that allow journals to game the system by overpublishing citable non-article outputs, such as editorials (Jain, et al., 2021; Lei et al., 2020; Seglen, 1997; Thelwall & Fairclough, 2015). Thus, despite the simplicity and intuitive appeal of JIF-like calculations, they should be interpreted cautiously.

Empirical research assessing whether academics in a field find JIFs to be credible vary between those that find broad acceptance (implicit in: Currie & Pandher, 2020) or rejection (e.g., Hurtado & Pinzón-Fuchs, 2021; Meese, et al., 2017). There are two separate issues: whether journals in a field can be credibly ranked and whether rankings produced by JIF-like calculations agree with expert rankings. Of course, academics are also frequently sceptical about expert-based journal rankings too (Bryce et al., 2020) and different expert rankings may disagree substantially (Meese et al., 2017) so there is no "gold standard" against which bibliometric journal rankings can be compared. The second issue has been repeated investigated and the answer varies over time for a field (Walters, 2017). Using the expert-based Australian journal strata, Elsevier's Source Normalised Impact per Paper (SNIP) correlated better than the JIF with human judgement in 27 field-based categories. The SNIP advantage may be its normalisation for field differences that makes it more appropriate in large categories containing multiple fields. In the 26 monodisciplinary broad categories checked, the correlations were close to zero in the arts and humanities (0.2), and weak in the social sciences (0.2-0.4) but stronger elsewhere (0.4-0.8), ignoring the multidisciplinary category (Haddawy et al., 2016). SNIP also correlates better than JIF with expert-based rankings of business and management journals (Mingers & Yang, 2017). Journal h-indexes also correlate moderately with human rankings in some fields (Mingers & Yang, 2017; Serenko & Bontis, 2021), perhaps because they combine quality and quantity components, with larger journals being more recognised. For instance, there are stronger associations between departmental h-indexes and REF scores in Biology (ranging from 0.71 to 0.79), Chemistry (0.71 to 0.83), Physics (0.44 to 0.59) and Sociology (0.53 to 0.62) than with institutional normalized citation impact (ranging from 0.37 to 0.67) (Mryglod et al., 2015).

An analysis of the correlation between peer review quality ratings and field/year normalised journal citation rates (SNIP) for 19,130 articles from REF2014 in 36 UoAs published in 2008 found Spearman correlation strengths being zero or negative in four UoAs: Classics (-0.8); Art and Design: History, Practice and Theory; Theology and Religious Studies (-0.1), Arts Area Studies (0). The strongest remaining correlations occurred for Clinical Medicine, Chemistry (all 0.5) and Biological Sciences (0.6) and Economics and Econometrics (0.7) (HEFCE, 2015, Table A18). Thus, as for citation counts, field/year normalised journal impact is an imperfect indicator of journal article quality in most academic fields, its value varies greatly between fields, and it is useless in some arts and humanities fields.

*Gender bias in academia, peer review and bibliometrics*

There is wide suspicion that sexism affects evaluations of the work of female academics because sexism is not yet eradicated from society and because women are underrepresented globally in senior roles (UNESCO, 2022) and for academic prizes (Meho, 2021). Many lists of highly cited scholars are also male dominated. Nevertheless, the extent of the impact of sexism on peer review scores and citation counts in academia is contested. There are many studies showing that female candidates are or are not discriminated against in evaluations of their research, with no clear outcome (Begeny, et al., 2020; Ceci & Williams, 2011). Moreover, overall career statistics and perhaps also prizes favour men because of shorter female career lengths (Huang et al., 2020). It is therefore possible that female-authored research is generally fairly judged in some fields but not others, such as those generating "chilly climates" for women (Biggs et al., 2018; Else, 2018; Overholtzer & Jalbert, 2021). Intersectional factors may well also be relevant, with women that are also from other disadvantaged groups being particularly affected in some or all fields (Banda, 2020; Wilkins-Yel et al., 2019).

Many studies have investigated whether female-authored papers tend to be less cited than male-authored papers, with the suspicion of direct citation sexism through men preferring to cite male authors in some or all fields (e.g., Wang, et al., 2021). Sexist citation practices may also be indirect, if the achievements of male authors are more celebrated, making their work more likely to be noticed and cited (Merton, 1968). Similarly, if men tend to cite their friends and these are men then this would generate a second order sexist citation bias against women. The empirical evidence for sexist citation is mixed, however, with the largest-scale evaluation with the most robust citation indicator suggesting a small female citation advantage in six out of seven large predominantly English-speaking countries (Thelwall, 2020). These national averages may hide individual fields where females are slightly less cited, however (e.g., Andersen et al., 2019; Maliniak, et al., 2013).

The strongest easily evidenced gender difference in academia is between fields rather than in citations. In many countries women numerically dominate some fields (e.g., nursing, allied health professions, veterinary science) and men numerically dominate others (e.g., mathematics, philosophy, physics, engineering) in terms of personnel (UNESCO, 2022) and publications (Thelwall et al., 2019; Thelwall et al., 2020). Africa may have the least gender variation between fields (at least in terms of students: UNESCO, 2022) and there is paradoxically greater gender inequality between fields in countries where there is less gender inequality overall (Stoet et al., 2018; Thelwall & Mas-Bleda, 2020). There is also a male/female gender differentiation within fields, with women more likely to engage in people-related topics, to use qualitative methods (Thelwall et al., 2019; Thelwall et al., 2020) and to have societal progress goals (Zhang et al., 2021). These factors all cause second order effects in bibliometric studies and perhaps also for peer review. Second order gender effects in bibliometrics are likely to occur because topics and fields have different citation rates. Thus, even for a set of researchers within a field, if one gender is more cited than the other then this could be because of differing research methods or specialties rather than sexism affecting choices of citations (e.g., Downes & Lancaster, 2019). It is impossible to fully differentiate between the two because all research is different and narrowing down to a specific enough topic to avoid the likelihood of topic or methods differences is likely to generate too few articles to statistically identify any gender difference, given that it is likely to be small (i.e., large samples are needed to detect small effect sizes).

*Difficulties evaluating interdisciplinary research*

Depending on how it is defined, interdisciplinary research combines theories, methods and/or personnel from multiple disciplines to address a common goal (Aboelela et al., 2007; Arnold et al., 2021; Wagner et al., 2011). For convenience, it is sometimes equated with the use of references from multiple fields (Van Noorden, 2015). For the REF, interdisciplinary research is effectively defined as research that needs the expertise of multiple UoA panels to evaluate (REF, 2019), and this is now it is operationalised in the current article. This does not directly match existing definitions of interdisciplinarity because it is evaluation-focused rather than input- or goal-focused, but it seems likely to have a large overlap in practice because both definition types involve multiple disciplines.

Interdisciplinary research is useful for applied research to address societal issues and for basic science that targets such issues as a longer-term goal (Gibbons et al., 1994; Stokes, 1997). Citation counts are likely to be less useful for evaluating interdisciplinary research than for single discipline research because its significance is more likely to be at least partly determined by non-academics judging its societal value (Gibbons et al., 1994; Whitley, 2000). Thus, factors unrelated to citations seem likely to be more important for interdisciplinary research quality judgements, and it is intrinsically complex to evaluate (Huutoniemi, 2010). There is no large-scale citation-based empirical evidence to support this claim, however.

Citation analyses of interdisciplinary research have tended to evaluate the extent to which average citation counts for interdisciplinary research relate to the average citation counts of the constituent fields. It has been shown, for example, that interdisciplinary research citation counts can tend to be greater or less than the average of the constituent fields, depending on the fields in question (Levitt & Thelwall, 2008). Using three dimensions of diversity (Stirling, 2007), combining a greater number of fields associates with more citations but combining dissimilar fields associates with fewer citations (Yegros-Yegros et al., 2015). Thus, from a citation analysis perspective the effect of interdisciplinarity is unclear and no hypotheses are implied for the overall relationship between interdisciplinarity and research quality.

# Methods

The research design was to apply bibliometrics to a set of articles with peer review scores and assess whether replacing peer review scores with bibliometric equivalents would introduce systematic score shifts suggestive of bias. Whilst the method is neutral about whether the changes would be introducing or removing bias, because the peer review scores might be biased, the former seems more likely.

*Data*

The data used in this analysis is 148,977 confidential provisional REF2021 scores from March 2021 for journal articles submitted by UK academics for assessment. These had to be first published between 2014 and 2020 to be in scope. Articles from the University of Wolverhampton were redacted for confidentiality reasons. The 148,977 scores are considered sensitive and had to be deleted by 9 May 2021, with only aggregates being subsequently published. REF2021 is split into 34 Units of Assessment (UoAs), each of which has a team of expert reviewers, predominantly senior UK academic researchers. Each article is allocated at least two primary reviewers who undertake to read it and agree a single quality score to encompass rigour, originality, and significance. The scores are 0 (unrated), 1* (recognised nationally), 2* (recognised internationally), 3* (internationally excellent), or 4* (world leading). The 318 unrated articles were removed because these sometimes indicated that an authorship claim had not been accepted rather than that the article was not of national quality. Many articles were submitted by multiple authors. Such duplicates were removed within UoAs or Main Panels, as appropriate for the aggregation level reported. When duplicates had different quality scores, the median was used, or a randomly selected median when there were two.

For the bibliometric data, the articles were matched against a copy of Scopus downloaded in January 2020 (to match when the bibliometric data for REF2021 would have been available). Articles were primarily matched with Scopus records by DOI (n=133,218), with a few extra matches (n=997) found by automatically comparing titles and using manual checks of the automatic matches. Only articles from 2014-17 were used (n=73,612, see also the first table below for exact numbers for the different experiments) to allow at least a three-year citation window (i.e., each article had at least three full years to attract citations), which should give a moderate correlation with long term citations in most fields (Wang, 2013).

Citation counts are not useful in this context because each UoA combines multiple years and fields (especially for interdisciplinary research). A field normalised log-transformed citation score (NLCS) was therefore calculated for each article as follows (Thelwall, 2017). First, all citation were replaced with the log transformation ln(1+x) to reduce skewing, diminishing the influence of individual highly-cited articles. Next, each log-transformed citation count was divided by the average for the narrow Scopus field and year it was in, giving the NLCS. Articles in multiple fields were instead divided by the average of the relevant field averages (averaging across all articles, not just UK articles). By design, NLCS are unbiased in the sense that they can fairly be compared between articles from different fields and years. An NLCS of 1 is always a world average score and higher values indicate more (log-transformed) citations than average for the field(s) and years of the article. There are up to 330 Scopus narrow fields in each year, so the field normalisation is relatively fine grained.

A journal impact indicator was also calculated for each journal as the mean of the NLCS of all articles in the journal for the given year. This is called here the journal mean NLCS, or JMNLCS. This is an average citation impact indicator for a journal. It is preferable to the well-known Journal Impact Factor because it adjusts for the skewed nature of citation counts (de Solla Price, 1976) and has a longer citation window.

Articles were grouped into HEIs according to the HEI that had submitted them. The gender analysis uses the first author gender of each article, as recorded in Scopus. First authors were assumed to be male or female if their first name, if recorded in Scopus, matched a first name that is used at least 90% for one gender in the UK, according to GenderAPI.com social media profiles or 1990 US census data (Larivière et al., 2013). Articles with authors having relatively gender-neutral names, such as Sam, were ignored for the gender analysis. Nonbinary genders were not detected because there was no practical non-intrusive way to identify them. Articles were regarded as interdisciplinary if they were flagged as such in the REF database, either by the submitting institution or the UoA panel members. This information was not assigned systematically and there were differences between HEIs in the extent to which these labels were used, so the quality of this data is weak.

### Analysis

A bibliometric calculation was carried out to mimic the REF procedure to assess how bibliometrics might impact the results for institutions.

For each HEI and UoA, the UK REF peer review stage calculates the number outputs rated 1*, 2*, 3*, or 4*. To closely mimic this process, the articles from each UoA were ranked in order from the lowest to the highest scoring on a bibliometric (i.e., NLCS or JMLNCS), and then thresholds set so that the same number of articles were in each category. In the case of ties, articles were arranged randomly so that the number of articles in each category was exact. Next, the average score of each HEI was calculated using (a) the peer review scores and (b) the bibliometric star scores. This calculation is sometimes informally called an institution's Grade Point Average (GPA). For each institution, the peer review GPA was subtracted from the bibliometric GPA to give the GPA increase (possibly negative) due to the switch to bibliometrics. Within each UoA institutional peer review GPAs were then correlated with GPA increases to assess for bias based on institutional research quality (as measured by peer review GPAs). A positive correlation would indicate that high GPA institutions gained from bibliometrics, and a negative correlation would indicate that they lost from it, whereas a zero correlation would indicate the lack of a (linear) bias. This correlation is not assessing the accuracy of the predictions but systematic factors behind changes.

In practice, bibliometrics in the REF were ignored in most UoAs and in the remaining 11 UoAs were probably used to arbitrate when two reviewers disagreed on an article (Wilsdon et al., 2015). It is not possible to simulate this application in the absence of information about which articles were difficult to evaluate. Such information was not made available to the project.

## Results

The number of articles 2014-17 and HEIs varies considerably between UoAs (Table 1). Overall, the number of articles and HEIs per UoA varies greatly, interdisciplinary research is rare, and men dominate Main Panel B UoAs.

Table 1. The number of articles, HEIs, first author male/female genders, and interdisciplinary articles analysed. All were submitted to REF2021 and matching a Scopus journal article 2014-17.

| # | UoA or main panel | HEIs | Female | Male | Interdisc | Monodisc | Articles |
|---|---|---|---|---|---|---|---|
| 1 | Clinical Medicine | 31 | 1948 | 2695 | 682 | 5289 | 5971 |
| 2 | Public Health, Health Services & Primary Care | 33 | 936 | 984 | 336 | 2043 | 2379 |
| 3 | Allied Health Professions, Dentistry, Nursing & Pharmacy | 89 | 2296 | 2124 | 850 | 5031 | 5881 |
| 4 | Psychology, Psychiatry & Neuroscience | 92 | 1997 | 2134 | 499 | 4496 | 4995 |
| 5 | Biological Sciences | 44 | 1252 | 1841 | 305 | 3535 | 3840 |
| 6 | Agriculture, Food & Veterinary Sciences | 25 | 610 | 686 | 203 | 1621 | 1824 |
| 7 | Earth Systems & Environmental Sciences | 40 | 519 | 1091 | 342 | 1936 | 2278 |
| 8 | Chemistry | 41 | 458 | 1088 | 294 | 1817 | 2111 |
| 9 | Physics | 44 | 282 | 1218 | 229 | 2913 | 3142 |
| 10 | Mathematical Sciences | 54 | 354 | 1909 | 314 | 2819 | 3133 |
| 11 | Computer Science & Informatics | 89 | 335 | 1615 | 423 | 2406 | 2829 |
| 12 | Engineering | 88 | 1128 | 4670 | 1330 | 9408 | 10738 |
| 13 | Architecture, Built Environment & Planning | 37 | 321 | 745 | 99 | 1345 | 1444 |
| 14 | Geography & Environmental Studies | 56 | 561 | 1005 | 105 | 1844 | 1949 |
| 15 | Archaeology | 24 | 105 | 139 | 22 | 276 | 298 |
| 16 | Economics & Econometrics | 25 | 140 | 718 | 25 | 997 | 1022 |
| 17 | Business & Management Studies | 107 | 1753 | 3771 | 361 | 6487 | 6848 |
| 18 | Law | 67 | 391 | 501 | 59 | 933 | 992 |
| 19 | Politics & International Studies | 56 | 422 | 856 | 84 | 1354 | 1438 |
| 20 | Social Work & Social Policy | 75 | 892 | 722 | 168 | 1674 | 1842 |
| 21 | Sociology | 37 | 376 | 380 | 142 | 708 | 850 |
| 22 | Anthropology & Development Studies | 22 | 193 | 231 | 29 | 489 | 518 |
| 23 | Education | 82 | 881 | 730 | 202 | 1670 | 1872 |
| 24 | Sport & Exercise Sciences, Leisure & Tourism | 60 | 404 | 862 | 154 | 1403 | 1557 |
| 25 | Area Studies | 20 | 107 | 120 | 49 | 214 | 263 |
| 26 | Modern Languages & Linguistics | 41 | 263 | 196 | 51 | 487 | 538 |
| 27 | English Language & Literature | 79 | 206 | 162 | 61 | 352 | 413 |
| 28 | History | 76 | 212 | 330 | 44 | 549 | 593 |
| 29 | Classics | 17 | 19 | 34 | 9 | 48 | 57 |
| 30 | Philosophy | 35 | 92 | 333 | 18 | 469 | 487 |
| 31 | Theology & Religious Studies | 22 | 23 | 57 | 16 | 74 | 90 |
| 32 | Art & Design: History, Practice & Theory | 69 | 225 | 230 | 58 | 530 | 588 |
| 33 | Music, Drama, Dance, Performing Arts, Film & Screen Studies | 68 | 120 | 148 | 46 | 258 | 304 |
| 34 | Communication, Cultural & Media Studies, Library & Information Man. | 54 | 220 | 244 | 57 | 471 | 528 |
| A | Main Panel A (UoAs 1-6) | 128 | 8168 | 9350 | 2519 | 19951 | 22470 |
| B | Main Panel B (UoAs 7-12) | 105 | 2973 | 11260 | 2778 | 20748 | 23526 |
| C | Main Panel C (UoAs 13-24) | 126 | 6294 | 10379 | 1395 | 18755 | 20150 |
| D | Main Panel D (UoAs 25-34) | 129 | 1477 | 1840 | 402 | 3430 | 3832 |

## *RQ1,2: Article level and journal-level citation-based indicators*

With one exception, HEIs with lower GPAs tend to gain from the bibliometric predictions. They always gain from JMNLCS predictions and usually gain from NLCS predictions, except in UoA 9. This tendency is moderate or strong in all UoAs except two (1 and 9). The correlations are statistically significantly different from 0 in all cases except UoA 1 (NLCS and JMNLCS), UoA 9 (NLCS), and UoA 25 (NLCS). Thus, journal-level and article-level bibliometrics disadvantage higher scoring institutions in almost all fields.

For the article-level NLCS, the most likely factor behind the result for lower numbered UoAs is that citations imperfectly reflect research quality so that weaker articles occasionally become highly cited, whereas stronger articles sometimes attract few citations. A department consistently producing high quality research could therefore expect to have some rarely cited articles and a department constantly producing lower quality research could expect to have some highly cited articles. Thus, whilst generally higher scoring departments tend to produce more cited research, they tend to be more consistent at producing high quality research than highly cited research. Chemistry is an exception. In this case there is a very weak tendency for the highest scoring departments to be more consistent at producing highly cited work than high quality work. It is possible that some departments selected articles partly on bibliometrics, in the knowledge that they may be consulted in REF evaluations. For higher numbered UoAs, where citations and impact factors have little correlation with research quality, the negative correlations are a statistical effect of replacing genuine scores with almost random noise and then averaging both. Thus, although the magnitude of the correlations are similar across all UoAs and the practical implications are the same (bias against higher scoring departments), there are at least two distinct causes.

For the journal-level JMNLCS, the above argument largely applies, except for the chemistry exception. Again, for lower numbered UoAs, departments tending to produce high quality research tend to be more consistent in producing high quality articles than in getting them published in high impact journals. Clinical Medicine is a partial exception, in that good departments are almost equally able to produce consistently high-quality research and publish in consistently high impact journals.

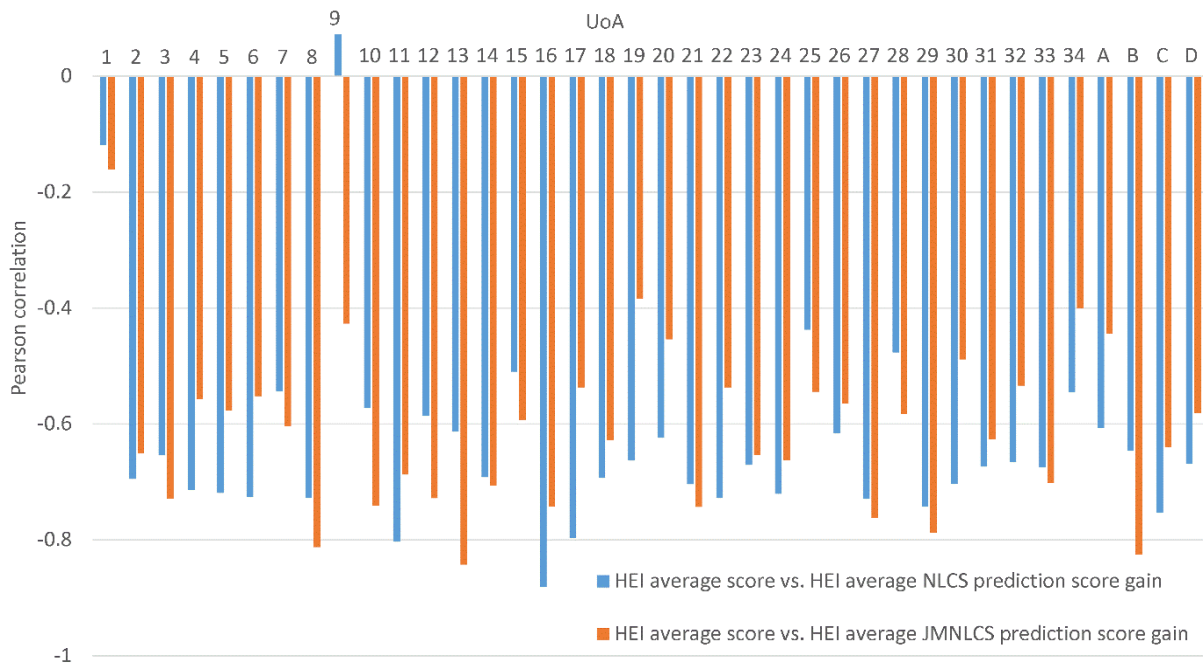

Figure 1. Pearson correlations between HEI average scores and HEI average prediction gains from allocating scores only with NLCS or JMNLCS.

## RQ3: Gender differences

There is some evidence of a weak tendency for female first-authored research to gain from bibliometric score allocation in some fields (Figure 2). The error bars include zero in almost all UoAs and the difference is marginal for the exceptions. It is not reasonable to draw strong conclusions for individual UoAs in this case because the marginal results are to be expected whenever many confidence intervals are drawn, even if there are no underlying differences (Rubin, 2017). Nevertheless, the female advantage is positive in 26 out of 34 UoAs for NLCS (p=0.001 for a post-hoc binomial test for gender difference α=0.5) and in 25 out of 34 UoAs for JMNLCS (p=0.001 for a post-hoc binomial test for gender difference α=0.5), giving statistical evidence of an overall female gain from bibliometrics. Moreover, the difference is statistically significant and positive for journal impact (JMLNLCS) in Main Panel B (mainly physical sciences and engineering). It is also statistically significant and positive for article citations (NLCS) in Main Panel C (mainly social sciences).

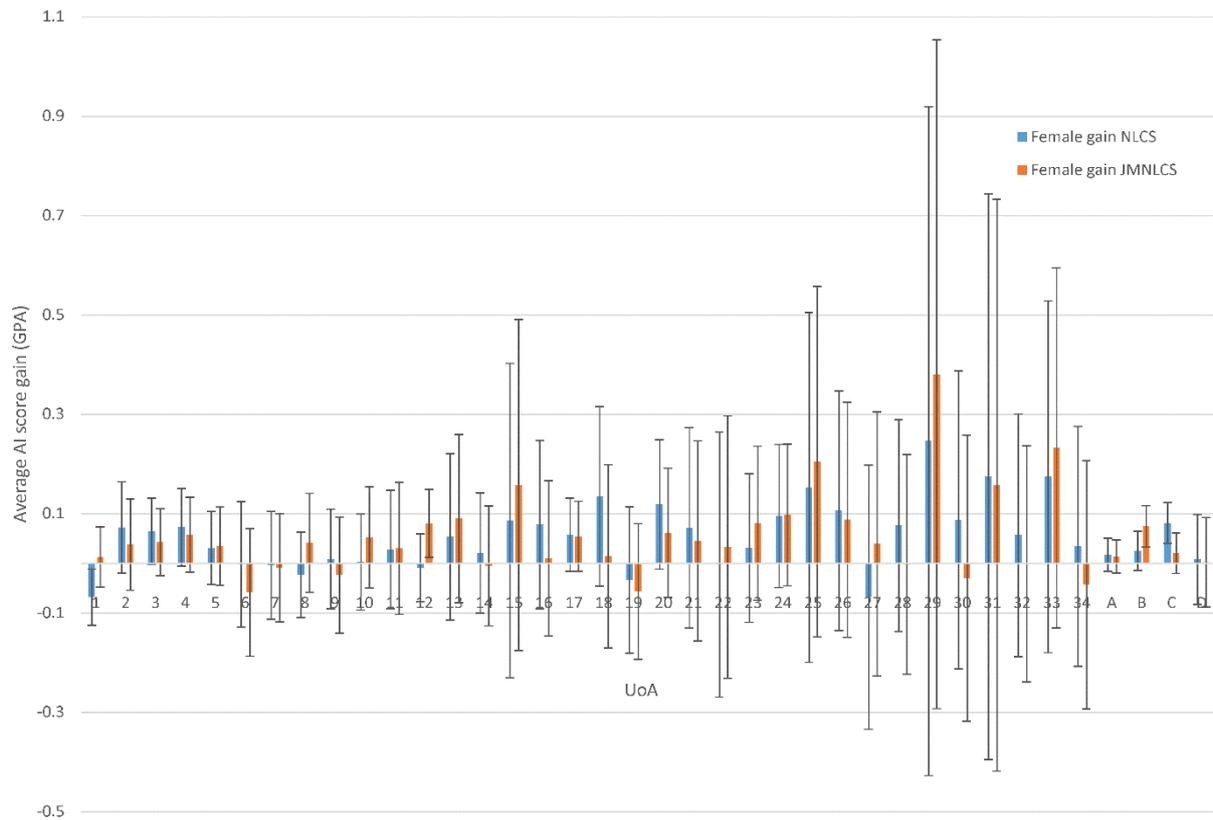

Figure 2. Prediction gains for female researchers compared to male researchers from allocating scores with NLCS or JMNLCS instead of peer review.

### *RQ3: Interdisciplinary research*

There is some evidence of a moderate tendency for interdisciplinary research to gain from bibliometric score allocation in some fields, but not overall (Figure 3). Whilst the error bars contain zero in most cases, UoAs 16 and 18 interdisciplinary research has statistically significant moderate advantages from both NLCS and JMNLCS. There is a weak but statistically significant bibliometric interdisciplinary gain for Main Panel B JMNLCS and Main Panel C NLCS. The overall UoA pattern is not statistically significantly different from equal, however, with 20 out of 34 UoAs having an advantage with NLCS (p=0.081 for a post-hoc binomial test for gender difference α=0.5) and 21 with JMNLCS (p=0.054 for a post-hoc binomial test for gender difference α=0.5).

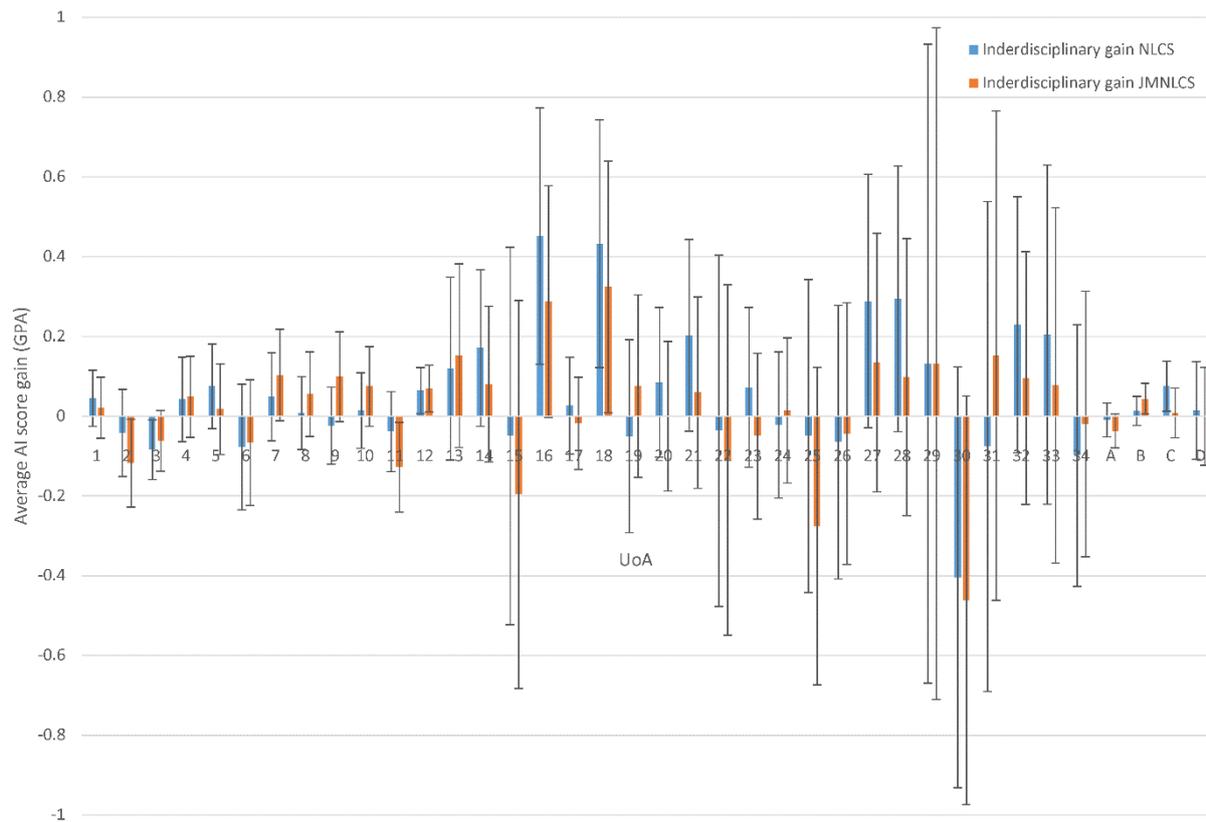

Figure 3. Score gains for interdisciplinary research compared to monodisciplinary research from allocating scores with NLCS or JMNLCS instead of peer review.

## Discussion

The analysis has many limitations. The data is from the UK, and bibliometrics may have a different value relative to peer review in other countries and for different peer review goals. Other field/year normalised indicators may have produced different results, particularly if they did not take citation skewing into account. The 34 UoAs used for the analysis are relatively coarse and a different categorisation scheme may have produced slightly different outcomes. The results may also change over time, and particularly for the journal-level analysis with the continued rise of megajournals. The gender detection may have introduced a second order bias related to ethnicity for names that were not detected with the algorithm. Perhaps most importantly, the interdisciplinary research flag may be inaccurate. It is possible that interdisciplinary differences found are second order effects from large strong or weak HEIs using it differently from average. Longer citation windows are also sometimes needed to assess interdisciplinary research citations (Chen et al., 2022), which may also have been a factor.

For RQ1 and RQ2, the results seem to be the first of their kind and so cannot be compared to prior findings. Nevertheless, whilst the bibliometric disadvantage for higher scoring departments has a simple and logical explanation, it does not seem to have been remarked on in previous studies or in evaluation criteria for national research evaluation exercises or considered in performance related funding procedures. These results are limited by the bibliometric ties being randomly allocated higher or lower scores. Whilst this simulates how the bibliometrics would have to be used if the exact number of articles in each star rating class is predetermined, such a use would be unrealistic in practice unless an assessment had fixed quotas for quality scores, such as to norm reference between fields in the assessment practice. Thus, this random assignment could be the reason why the bibliometrics have a damping effect in most UoAs. Nevertheless, the problem of ties would need to be resolved somehow if bibliometrics were to be used, and there does not seem to be a fairer solution.

The minor gender bias in favour of women from both article-level and journal-level citations aligns with prior research of small gender citation advantage of women compared to men in the UK (Thelwall, 2020). It extends this by suggesting that citations slightly overestimate the quality of female first-authored research. This conflicts with a previous suggestion that citations underestimate the significance of female first authored research because it tends to be more read than cited (Thelwall, 2018). Thus, either the peer review scores have a slight male bias, such as by insufficiently considering wider societal value, or the previous argument based on readership information (Thelwall, 2018) was incorrect, perhaps because it did not consider non-educational impacts, such as commercial value.

The lack of an overall trend in the relationship between interdisciplinarity and any citation advantage is the first result of its kind but aligns with prior arguments that interdisciplinarity is complex, with no simple quality pattern (Huutoniemi, 2010) including for its citation relationship (Yegros-Yegros et al., 2015). The existence of exceptions in relatively small UoAs may be due to relatively stable interdisciplinary fields, such as econophysics, with high levels of citation (Sharma & Khurana, 2021).

# Conclusion

This study found that departments producing better research tend to be disadvantaged when bibliometrics are used, even in fields where they have high correlations with quality scores. This may be due to the damping effect of randomly assigning tied bibliometric scores to higher or lower classes, however. A practical implication of this that evaluation exercises relying on bibliometrics should be aware of this potential deficiency and either accept it or take steps to remedy it. This applies equally for exercises, like the REF, where bibliometrics are used to support peer review rather than to replace it. For example, if the bibliometric information does not help make a quality decision in cases where REF peer reviewers disagree, it would be logical to favour a quality score that aligned with the departmental average. This would give a small nudge to partly offset the bibliometric bias.

The minor gender advantage for females compared to males for bibliometrics in the UK should be reassuring for those seeking to use bibliometrics to support research assessment in the sense that it is unlikely to introduce a bias against women. Given the additional obstacles faced by women in society and academia, a small citation bias in their favour seems unproblematic.

The results also suggest that interdisciplinary research is not disadvantaged overall by bibliometrics. Nevertheless, evaluators should nevertheless be watchful for individual high or low citation interdisciplinary fields in which bibliometrics may be misleading.

# Acknowledgement

This study was funded by Research England, Scottish Funding Council, Higher Education Funding Council for Wales, and Department for the Economy, Northern Ireland as part of the Future Research Assessment Programme (https://www.jisc.ac.uk/future-research-assessment-programme). The funders had no role in the design or execution of this study. The content is solely the responsibility of the authors and does not necessarily represent the official views of the funders.